\documentclass[journal]{IEEEtran}
\usepackage{graphicx}
\usepackage{siunitx}

\usepackage{amsthm,amsmath,amssymb}
\usepackage{mathrsfs}
\usepackage{bm}
\usepackage{booktabs}
\usepackage{multirow}

\DeclareMathOperator*{\argmin}{arg\,min}

\usepackage{pifont}
\newcommand{\cmark}{\ding{51}}%
\newcommand{\xmark}{\ding{55}}%
\ifCLASSINFOpdf
\else
\fi
\begin{document}
%
\title{\huge{Data and Physics Driven Learning Models for Fast MRI} \\ \Large{Fundamentals and Methodologies from CNN, GAN to Attention and Transformers}}
%
%
%

\author{Jiahao Huang,
        Yingying Fang,
        Yang Nan,
        Huanjun Wu,
        Yinzhe Wu,
        Zhifan Gao,
        Yang Li, \\
        Zidong Wang,~\IEEEmembership{Fellow,~IEEE,}
        Pietro Lio,~\IEEEmembership{Fellow,~IEEE,}
        Daniel Rueckert,~\IEEEmembership{Fellow,~IEEE,} \\
        Yonina C. Eldar,~\IEEEmembership{Fellow,~IEEE,}
        and Guang Yang,~\IEEEmembership{Senior Member,~IEEE}
\thanks{Jiahao Huang, Yingying Fang, Yang Nan, Huanjun Wu, Yinzhe Wu and Guang Yang are with the National Heart and Lung Institute, Imperial College London, Guy Scadding Building, Dovehouse St, London, SW3 6LY, UK. (Send correspondence to: g.yang@imperial.ac.uk).}
\thanks{Huanjun Wu and Yinzhe Wu are also with the Department of Bioengineering, Imperial College London, South Kensington, London, SW7 2AZ, UK.}
\thanks{Zhifan Gao is with the School of Biomedical Engineering, Sun Yat-sen University, Guangzhou, Guangdong, 510006, China.}
\thanks{Yang Li is with the School of Automation Sciences and Electrical Engineering, Beihang University, Beijing, China.}
\thanks{Zidong Wang is with the Department of Computer Science, Brunel University London, London, UB8 3PH, UK.}
\thanks{Pietro Lio is with the Computer Laboratory, University of Cambridge, Cambridge, CB3 0FD, UK.}
\thanks{Daniel Rueckert is with the Department of Computing, Imperial College London, South Kensington Campus London, SW7 2AZ, UK. and Technical University of Munich, Munich, 80333, Germany.}
\thanks{Yonina C. Eldar is with Faculty of Mathematics and Computer Science Weizmann Institute of Science, 7610001, Rehovot, Israel.}
\thanks{This study was supported in part by the BHF (TG/18/5/34111, PG/16/78/32402), the ERC IMI (101005122), the H2020 (952172), the MRC (MC/PC/21013), the Imperial-NTU Research Collaboration Fund, and the UKRI Future Leaders Fellowship (MR/V023799/1).}
\thanks{Manuscript received December 01, 2021.}}

%
%

\markboth{IEEE Signal Processing Magazine, August~2022}%
{Huang \MakeLowercase{\textit{\textit{et~al.}}}: Data and Physics Driven Learning Models for Fast MRI}
%



\maketitle

\begin{abstract}
Research studies have shown no qualms about using data driven deep learning models for downstream tasks in medical image analysis, e.g., anatomy segmentation and lesion detection, disease diagnosis and prognosis, and treatment planning. However, deep learning models are not the sovereign remedy for medical image analysis when the upstream imaging is not being conducted properly (with artefacts). This has been manifested in MRI studies, where the scanning is typically slow, prone to motion artefacts, with a relatively low signal to noise ratio, and poor spatial and/or temporal resolution. Recent studies have witnessed substantial growth in the development of deep learning techniques for propelling fast MRI. This article aims to (1) introduce the deep learning based data driven techniques for fast MRI including convolutional neural network and generative adversarial network based methods, (2) survey the attention and transformer based models for speeding up MRI reconstruction, and (3) detail the research in coupling physics and data driven models for MRI acceleration. Finally, we will demonstrate through a few clinical applications, explain the importance of data harmonisation and explainable models for such fast MRI techniques in multicentre and multi-scanner studies, and discuss common pitfalls in current research and recommendations for future research directions.
\end{abstract}


%
\IEEEpeerreviewmaketitle

\section{Introduction}
%
%
%
%
\IEEEPARstart{A}{lthough} Magnetic Resonance Imaging (MRI) is one of the most sensitive yet suitable technologies for providing crucial measurements in clinical diagnosis, prognosis and treatment planning, it is limited in use due to prolonged scanning time, which can cause significant expense with patient comfort and compliance concerns. After selecting an image field of view and spatial/temporal resolution, the minimum acquisition time is usually determined by the number of raw data required to fulfil the Nyquist requirements in the acquisition space, namely \textit{k}-space~\cite{hollingsworth2015reducing}.

MRI technology has evolved from linear analytic reconstruction to nonlinear iterative reconstruction methods during the last several decades. Because of their simplicity, linear analytic reconstruction approaches have been widely employed in commercialised scanners. However, because no scanning object prior is considered in these linear analytic reconstruction methods, these approaches can suffer from relatively large residual artefacts~\cite{wang2021deep}. 

There has been a great interest in developing and employing compressed sensing (CS) theory and applications in MRI to minimise acquisition duration. According to CS theory, if our target MRI images are sparse, with signal information in only a small percentage of pixels (e.g., in MR angiogram), or if the image can be mathematically transformed into a space with certain sparsity, we can use that sparsity to recover a high quality image from significantly less acquired data. However, because of the nonlinear iterative nature of CS based methods, the reconstruction procedure is significantly slowed, making typical CS-MRI difficult for use in clinical settings~\cite{wang2021deep}.

Deep learning is a technology that employs neural networks with deep layers of processing units to discover complicated patterns in vast amounts of data, which is driving today's artificial intelligence (AI) growth. In general, deep neural networks learn underlying characteristics and crucial basis functions as non-linear representations of data that best match the task for which they are trained. As part of an optimisation process to discover the greatest potential characteristics, it does so without human intervention. Given more data, unlike traditional statistical and machine learning approaches, the performance of deep learning models does not plateau but instead rises according to a power law. AI and deep learning have long been used in medical imaging. For decades, machine learning has been utilised in radiography for computer-aided diagnosis and detection. Recent improvements in computer vision---specifically, convolutional neural networks---have already been leading to promising achievements in a variety of fields of medical imaging, including disease diagnosis and classification, anatomy segmentation and lesion detection, as well as medical image synthesis~\cite{shen2017deep}. Deep learning has been used in computational MRI as inspired by those successful medical image analysis applications, and it has demonstrated the capability to dramatically speed up MR reconstruction. Deep learning, unlike CS and other iterative reconstruction algorithms, avoids difficult parameter adjustment for optimisation and conducts ultra-fast online reconstruction with the help of offline training on massive volumes of data~\cite{chandra2021deep}.

Deep learning based fast MRI algorithms are roughly divided into two categories, i.e., those using unrolling and those not~\cite{liang2020deep}. Unrolling-based techniques begin with a posed optimisation problem, the answer to which is the image to be recovered. Then, they unroll an iterative optimisation process using a deep network. As a result, the network architecture of an unrolling based technique is built on the steps that arise from iterations. Network training is used to learn the parameters and functionalities of the reconstruction models. The approaches that are not dependent on unrolling directly, on the other hand, employ deep neural network designs developed for problems other than reconstruction to learn the mapping from input to output and may also include some domain knowledge in MRI into the network training.

Typical deep learning models for fast MRI embody convolutional neural networks (CNNs) and generative adversarial networks (GANs), which have shown promising results. Recurrent neural networks (RNNs) have also been employed for solving spatial and temporal dependencies of MRI reconstruction when dealing with high dimensional data recovery. More recently, attention mechanisms and transformers are in development for MRI acceleration. 

The main goal of this article is to provide an overview of deep learning based data driven algorithms in fast MRI by highlighting their distinct qualities as well as their commonalities. We make an effort not only to summarise the contents scattered across the literature but also to analyse the links between these approaches. This article by no means intends to include an exhaustive list of references (cf. topical reviews by~\cite{WangGe2019,knoll2020deep,wen2020transform,liang2020deep,wang2021deep}) from all contributions in deep learning based fast MRI as the area is rapidly growing (cf. systematic review and meta-analysis by~\cite{chen2022ai}). Rather the selected approaches reviewed here will serve as representative examples and a comprehensive tutorial for understanding the concept and importance of deep learning powered and data and physics driven fast MRI.

\section{MRI Basics and Conventional Acceleration}

\subsection{Lay Summary of MRI}

Magnetic Resonance Imaging (MRI) uses magnets and radio waves to picture inside the human body without imposing harmful radiation. It provides excellent anatomical details and can be used to aid in disease diagnosis and surgical guidance as well as research into normal biological function. The MRI scanner is a giant circular magnet that employs a strong magnetic field---so strong that it can lift the weight of a double-decker bus. The MRI scanner creates signals that can be interpreted by a computer with the help of a computer-controlled radiofrequency (RF) system and magnetic field modulation to form pictures across the subject, such as a human patient.

Our bodies are made up of 60\% water on average, and the water molecules in our bodies are continually rotating in a random pattern. When these water molecules approach close to a magnet as powerful as the MRI scanner, however, they modify how they rotate, which the MRI scanner can detect. More specifically, MRI works because the nucleus at the centre of a hydrogen atom---a component of the water molecule---has magnetic qualities that lead all of the nuclei to become very weakly magnetised when exposed to a strong magnetic field. The hydrogen nuclei are able to absorb magnetic energy provided by a short RF pulse, but this magnetisation is too faint to detect conventionally. 

This RF pulse can also force the previously undetectable magnetisation away from the main field direction, and all of the individual nuclear magnets start to rotate (i.e., to precess) around the main field direction, much like a spinning top or a gyroscope when pushed away from the vertical. The spinning of the nuclear magnets will create a voltage in a receiver coil positioned around the scanned subject when the coil is set to the matched frequency. An analogy is thinking of running an electric motor in reverse that the coils rotate while the magnet remains stationary or the dynamo on a bicycle with the wheel motion geared down to rotate a magnet in a coil and create a current to power the light. The nuclear magnets create a high-frequency voltage in the coil (63.9MHz in a 1.5T standard clinical MRI scanner), which is amplified, digitised, and supplied into a computer. The scanner will be able to identify the presence of water in the body as a result of this.

By further adjusting the RF pulse and applying gradients to the main magnetic field, spatial information can be retrieved in a predictable manner throughout the scanned subject. Varied types of tissues in our body have different quantities of water. For example, bone has very little water, fat contains somewhat more, and blood contains a lot. This is why the different tissues on an MRI scan appear differently. Variations in water content and the rate at which nuclear magnetism decays after return back to its initial status provide contrast differences in the picture, but other processes can also be used to highlight blood flow and particular pathological characteristics.

\subsection{MRI Physics}

The magnetisation properties of atomic nuclei are used in MRI. In particular, nuclei with an odd number of protons, which have the angular momentum, are MR-active. In MRI research studies, $^{1}$H (Hydrogen), $^{13}$C (Carbon), $^{17}$O (Oxygen), $^{19}$F (Fluorine), and $^{31}$P (Phosphorus) are used; however, the isotope of hydrogen called protium is the most commonly used MR-active nucleus in clinical MRI. 

Protons that are ordinarily randomly orientated inside the water nuclei of the tissue being investigated are aligned using a strong and uniform external magnetic field $\mathcal{B}_0$. A typical clinical MR system has a $\mathcal{B}_0=1.5$T (Tesla), 1T=10,000G (gauss) compared to the magnetic field of the earth is about 0.5G. When a patient is exposed to $\mathcal{B}_0$, a prevalent misconception is that the hydrogen nucleus aligns with the external magnetic field. Actually, the magnetic moments of hydrogen nuclei, not hydrogen nuclei themselves, align with $\mathcal{B}_0$. The hydrogen nucleus spins on its axis rather than changing orientation.

\begin{figure*}[thpb]
  \centering
  \includegraphics[scale=0.55]{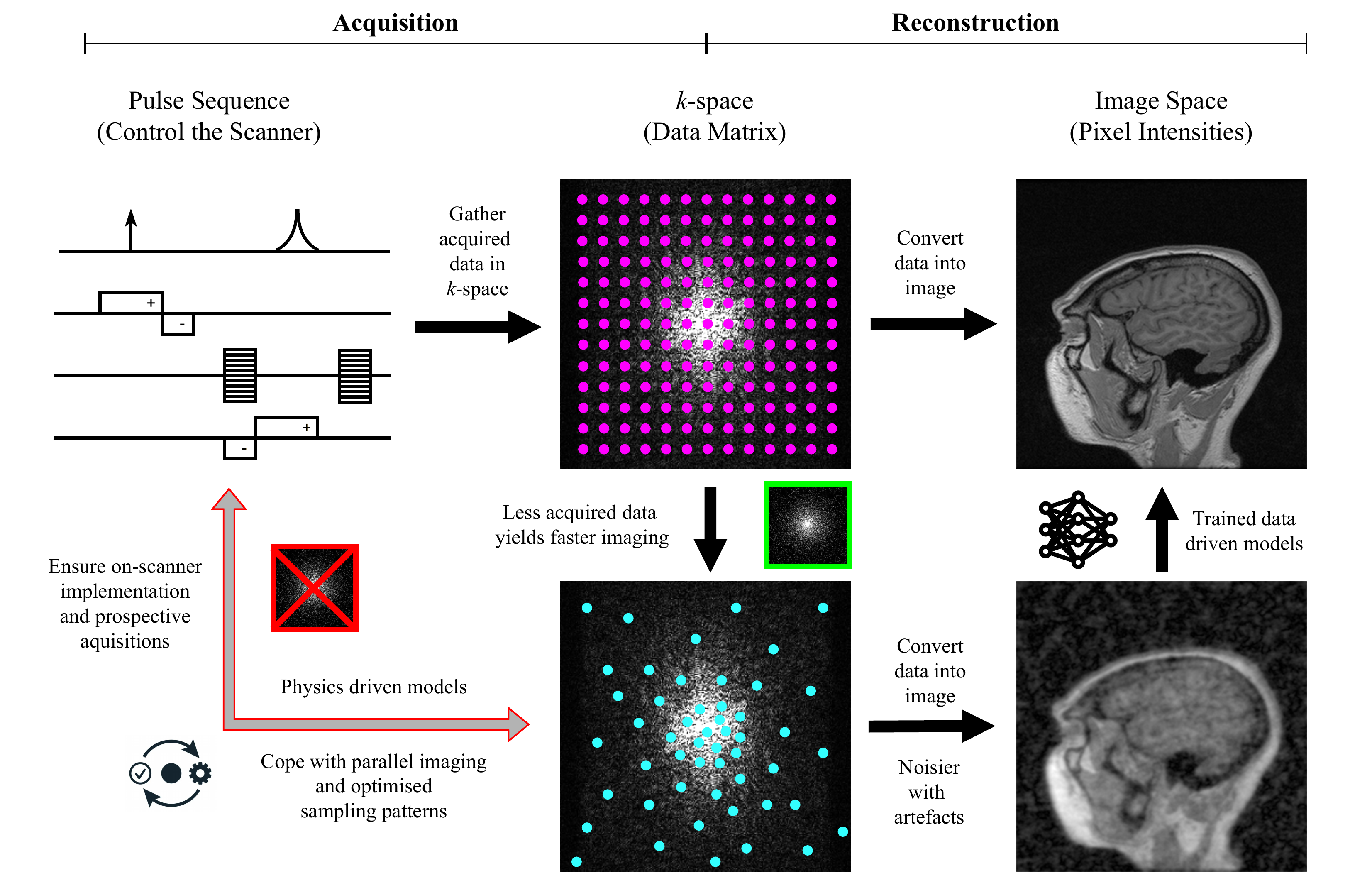}
  \caption{
    (Top) Image formation in MRI. The MRI scanner's actions and timing are dictated by a pulse sequence (Top Left). The collected data are put into a data matrix known as \textit{k}-space (Top Middle). When the data matrix is full, a procedure known as reconstruction is used to transform it into a picture (Top Right). If partial data are acquired (Bottom Middle), noisier and artefacts contaminated images are reconstructed (Bottom Right). Data driven models can remove the artefacts and noise, but not all data driven models without MR physics constraints can be implemented on the MRI scanner. For example, the green box shows a 2D Gaussian downsampling, but it is not feasible for on-scanner implementation. Physics driven models can cope with parallel imaging and optimised sampling patterns. 
  }
  \label{fig:1}
\end{figure*}

In general, MRI consists of 4 major steps to obtain the images, including excitation, relaxation, reception, and reconstruction (Figure \ref{fig:1}). 1) Excitation: the introduction of an external RF energy perturbs or disrupts this alignment (or magnetisation); 2) Relaxation: RF energy is emitted as the nuclei return to their resting alignment through different relaxation processes; 3) Reception: the emitted signals are monitored after a specific amount of time has passed from the first RF; 4) Reconstruction: the Fourier transformation is used to convert the signal's frequency information from each place in the scanned plane to matching intensity levels, which are then represented as grey level pixels. It is of note that the frequency of the applied RF pulses must match the proton's resonance frequency, i.e., Larmor frequency, causing it to tip out of alignment with the external magnetic field and precess around it. In other words, only when an oscillating magnetic field with a frequency closely match to the Larmor frequency (i.e., $\mathcal{B}_1$ field) is applied can the nuclear spin orientations be perturbed. The magnetisation at the microscopic level representing crowd behaviour of protons can be expressed as
\begin{equation}
\begin{aligned}\label{formula:larmor}
\mathsf{f}=\gamma\mathcal{B}_0,
\end{aligned}
\end{equation}
in which, $\mathsf{f}$ is the Larmor frequency and $\gamma$ is the gyromagnetic ratio. Therefore, at 1.5T, the Larmor frequency is 63.9 MHz = 42.6MHz/T (gyromagnetic ratio of hydrogen)$\times$1.5T (magnetic field strength). The RF system generates the $\mathcal{B}_1$ field at this frequency for a brief period of time (an RF pulse) to tip the magnetisation away from the $\mathcal{B}_0$ direction, resulting in transverse magnetisation detectable by an RF receive coil set to the same frequency. To achieve homogeneous excitation, a spatially uniform $\mathcal{B}_1$ field is required.

Different types of images can be created by varying the sequence of RF pulses applied and collected. The amount of time between successive pulse sequences applied to the same slice is known as the Repetition Time (TR). The delay between the delivery of the RF pulse and the reception of the echo signal is known as the Echo Time (TE). Tissue can be classified into two types based on its relaxation times: the spin–lattice relaxation time (T1) and the spin–spin relaxation time (T2). The time constant T1 (longitudinal relaxation time) controls the pace at which stimulated protons to return to equilibrium. It is the time it takes for spinning protons to realign themselves with the external magnetic field. The time constant T2 (transverse relaxation time) governs the pace at which excited protons attain equilibrium or move out of phase with one another. It's the time it takes for spinning protons to lose phase coherence with nuclei spinning perpendicular to the main field.

\subsection{Mathematics in MRI}

\subsubsection{Forward Model} In MRI, data is acquired by employing a succession of RF pulses and magnetic field gradients, which are characterised collectively by a pulse sequence (Figure \ref{fig:1}). A pulse sequence is a series of instructions that tells the MR scanner when and how strong to administer RF pulses, turn magnetic field gradients on and off, and turn on receivers to acquire the signal. An over-simplified continuous notation~\cite{doneva2020mathematical} of the measurement model is
\begin{equation}
\begin{aligned}\label{acquisition_c}
y(t) = \int x(\vec{r}) \exp^{-i 2 \pi \vec{\kappa}(t) \cdot \vec{r}} d \vec{r} = \mathfrak{F} \vec{\kappa}(t),
\end{aligned}
\end{equation}
in which $\vec{r}$ represents the spatial coordinates, $\vec{\kappa}(t)$ is the \textit{k}-space trajectory of the MR pulse sequence and $\mathfrak{F} \vec{\kappa}$ is the Fourier transform of $x(\vec{r})$. In reality, only a limited number of \textit{k}-space samples can be obtained, resulting in the discretised formulation, that is
\begin{equation}
\begin{aligned}\label{acquisition_d}
y(\vec{\kappa}_i) = \sum_j x(\vec{r}_j) \exp^{-i 2 \pi \vec{\kappa}_i \cdot \vec{r}_j}.
\end{aligned}
\end{equation}

\subsubsection{Inverse Model} In general, MRI reconstruction including fast MRI can be described as an \emph{inverse problem} mathematically. An inverse issue has no formal mathematical description, but from an applied standpoint, such problems are concerned with deriving causes from intended or observed consequences. This is often formalised as solving an operator equation, that is
\begin{equation}
\begin{aligned}\label{forwad_problem}
y = \mathcal{A}(x_\mathrm{true}) + n,
\end{aligned}
\end{equation}
in which $\mathcal{A}$ is a forward operator that maps model parameter $x_\mathrm{true} \in X$ to measured data $y \in Y$, i.e., $\mathcal{A}: X \to Y$, with observation noise $n$. 

In MRI, because the observed data can be considered as samples of the Fourier transform of the ideal signal, the MRI image reconstruction entails an inverse problem of Equation~\eqref{forwad_problem}, with the forward operator $\mathcal{A}$ represented as a discrete sampling operator concatenated with the Fourier transform $\mathfrak{F}$. Considering the complex-valued nature of MRI data (e.g., raw data acquired in the \textit{k}-space), noise $n$ is normally distributed, and the noise model of $\mathfrak{F}^{-1}(y)$ is Rician.  

As of MRI, image reconstruction can be formulated as an unconstrained optimisation to minimise the difference between the forward model and the measurements, that is
\begin{equation}
\begin{aligned}\label{unconstrained}
\hat{x} = \argmin_x || \mathfrak{M}(x) - y ||_2^2,
\end{aligned}
\end{equation}
in which $\mathfrak{M}(\cdot)$ represents the signal model. $x \in \mathbb{C}^N$ is the MRI images to be reconstructed, and $y \in \mathbb{C}^M$ is the MRI raw data measurement or acquisition. In theory, Equation~\eqref{unconstrained} is a convex optimisation problem that can be solved iteratively using conjugate gradient type methods and other accelerated and discretised algorithms, e.g., generalised minimal residual method. It is of note that because of the discretised and limited sampling, as well as the existence of noise, this reconstruction problem is ill-posed even with the simple Fourier model and a fully-sampled \textit{k}-space.

Classical Tikhonov regularisation is arguably the most prominent method for solving this ill-posed inverse problem by minimising the measure of data misfit that is penalised with a regulariser $\mathfrak{C}(x)$, which can be represented as a constrained optimisation
\begin{equation}
\begin{aligned}\label{inverse_problem}
\mathcal{R}_\zeta(y) := \bigg\{ \argmin_{x \in X} \frac{1}{2}|| \mathfrak{M}(x) - y ||_2^2 + \zeta \mathfrak{C}(x) \bigg\},
\end{aligned}
\end{equation}
in which the reconstruction operator $\mathcal{R}: X \to Y$ is a mapping that gives a point estimate $\hat{x}$ as the solution to Equation~\eqref{forwad_problem}. Here $X$ and $Y$ are Hilbert spaces (i.e., both are typically $L^2$ spaces). Therefore, classic Tikhonov ($L^2$) regularisation uses Hilbert-space norms $\mathfrak{C}(x) = \frac{1}{2} || y ||_2^2 $, which is the most commonly used penalty term, to regularise the inverse problems as Equation~\eqref{inverse_problem}. In addition, the data fidelity is ensured by $|| \mathfrak{M}(x) - y ||_2^2$ and the regularisation term $\mathfrak{C}(x)$ is introduced to impose prior knowledge with $\zeta$ as the corresponding trade-off between the regularisation term and the data fidelity term. 

Total variation (TV) based regularisation is widely used due to its edge-preserving and noise-suppression properties that is given as
\begin{equation}
\begin{aligned}\label{TV}
\mathfrak{C}_{\mathrm{TV}}(x) := || \nabla x ||_1,
\end{aligned}
\end{equation}
in which $\nabla$ is the spatial gradient and $|| \cdot ||_1$ is the L1 norm. A disadvantage of applying the TV regularisation approach becomes obvious when the $x_\mathrm{true}$ has more intricate, higher-order structures, such as piecewise linear regions, in addition to constant areas. In this situation, the TV may result in staircasing artefacts, which could be mitigated by using total generalised variation with higher-order derivatives.


\subsubsection{Fast MRI} The Nyquist-Shannon sampling criteria are used to determine the needed \textit{k}-space raw data once the intended field-of-view (FOV) and spatial resolution of the MRI images are specified. The sampling interval ($\Delta k = 1/FOV$) determines the measured FOV, while the highest sampled spatial frequency ($\Delta x = 1/2k_{\text{max}}$) determines the voxel size. Undersampling \textit{k}-space is one possible fast MRI method, resulting in an acceleration rate proportional to the undersampling ratio if pre-scans with calibrations are neglected. Early fast MRI approaches explored obtaining several lines in \textit{k}-space from a single RF stimulation by adopting multiple RF or gradient refocusing. These acceleration approaches are nevertheless classified as fully-sampled methods since they achieve the complete \textit{k}-space coverage required by the Nyquist-Shannon sampling requirement.

Partial Fourier imaging (PFI) is an undersampled approach based on the premise that the Fourier transform of a purely real function has complex conjugate symmetry in \textit{k}-space, hence only half of the \textit{k}-space in the phase encoding direction is required. In reality, however, more than half of the phase encoding is captured to give a strong phase correction, therefore the acceleration factor employing PFI is restricted to 2 and is associated with a loss in signal to noise ratio (SNR). 

Parallel imaging, on the other hand, is a fast MRI approach that employs many independent receiver channels. The acceleration factor of parallel imaging is restricted by the number and layout of receiver coils, which may produce certain image artefacts and raise the MRI scanner's manufacturing cost.

Iterative techniques are often employed to address the underdetermined inverse issue for MRI reconstruction when using sub-Nyquist sampling. CS is one of the ground-breaking methods for reconstructing from sub-Nyquist-sampled data. The technique solves the underlying restricted optimization issue by utilising some previous models, such as sparsity and low-rankness. The imaging model of the sub-Nyquist data may be stated in general as
\begin{equation}
\begin{aligned}\label{CS}
\mathcal{R}_\zeta(y) := \bigg\{ \argmin_{x \in X} \frac{1}{2}|| \mathfrak{M}(x) - y ||_2^2 + \zeta \mathfrak{R}(\mathfrak{D}x) \bigg\},
\end{aligned}
\end{equation}
in which $\mathfrak{M}=\mathfrak{U}\mathfrak{F}$,  $\mathfrak{U} \in  \mathbb{C}^{M \times N} (M\ll N) $ is an under-sampling matrix, $\mathfrak{F} \in  \mathbb{C}^{N \times N}$ is a Fourier transform matrix, $\mathfrak{D}$ is the sparsifying transformation and $\mathfrak{R}(\cdot)$ is a regularisation function to impose the sparsity of the data, e.g., $|| \cdot ||_1$ and $|| \cdot ||_{1-2}$ 
If a signal is sparse or compressible, it can be accurately reconstructed from a limited number of incoherent observations, according to CS theory. In the context of MRI, incoherent sampling can be accomplished by variable-density nonuniform sampling on a Cartesian grid or using non-Cartesian \textit{k}-space sampling. CS provides an intriguing relationship between data acquisition and visual information content: the sparser the image, the less information we need to gather. By using a suitable transform $\mathfrak{D}$, most MR images can be sparsified. By integrating the sparsity reserving functions $\mathfrak{R}$, this transformed sparsity prior is integrated into the reconstruction as a regularisation. Discrete Cosine Transform, wavelets and finite differences are popular $\mathfrak{D}$ options for MRI.




\section{Overview of Deep Learning Based Models}

In this section, we will provide an overview of the most widely used CNN, GAN and more advanced attention and transformer based models and in particular those for solving inverse problems.

\subsection{Basics of the CNN}

A neural network is a design motivated by the system of neurons in the human brain. A layer of a generic neural network can be generally formulated as:
\begin{equation}
\begin{aligned}\label{formula:cnn1}
y^{(l)}=\phi\left(\sum_{k=1}^{d} w^{(l)}_k x^{(l)}_k+b^{(l)}\right),   
\end{aligned}
\end{equation}
in which the output $y^{(l)}$ is a weighted, biased summation of the input $x^{(l)}$ acted by the activation function $\phi$.

Convolutional neural networks (CNNs) are a specialised kind of neural network whose layer is implemented by an operation same as convolution but without flipping the kernel:
\begin{equation} 
\begin{aligned}\label{formula:cnn2}
y_{i, j}^{(l)}=(x^{(l)} * k^{(l)})_{i, j}=\sum_{m} \sum_{n} x^{(l)}_{i+m, j+n}k^{(l)}_{m, n},
\end{aligned}
\end{equation}
in which the output $y_{i, j}^{(l)}$ is weighted sum of the input image $x$ by a filter $k$ of size $m$ by $n$.

Applying the convolution with filters of a certain size, the network layer is able to find certain features by looking at a smaller portion of the image. As a powerful feature extractor, CNN has gained much development including well-known VGG~\cite{Simonyan2015vgg} and ResNet architecture~\cite{he2016deep} to obtain a deeper network and larger perception and achieved tremendous success in a wide range of real applications.

\subsection{Basics of the GAN}

Generative adversarial network (GAN)~\cite{goodfellow2014generative} was first proposed by Goodfellow \textit{et~al.} in 2005.
In general, a GAN-based model reaches the generative model for the target via simultaneously training a generator $G_{\theta_G}(\cdot)$ which produces data to be distributed as real as possible to the real distribution $p_{\mathrm{data}}(x)$, and a discriminator $D_{\theta_D}(\cdot)$ that is capable of distinguishing the true data $x$ from the fake data $G_{\theta_G}(z)$. 
This adversarial training process can be described as a min-max game given as:
\begin{equation}
\begin{aligned}\label{formula:ganloss}
\mathop{\text{min}}\limits_{\theta_G} 
\mathop{\text{max}}\limits_{\theta_D}
\mathcal{L}(\theta_G, &\theta_D)
=\mathbb{E}_{x \sim p_{\mathrm{data}}(x)}
[\mathop{\text{log}} D_{\theta_D}(x)]\\
&+\mathbb{E}_{z\sim p_{z}(z)}
[\mathop{\text{log}} (1-D_{\theta_D}(G_{\theta_G}(z)))],
\end{aligned}
\end{equation}
in which $z$ is a random input of the generator sampled from a fixed latent distribution $p_{z}$, $\theta_G$ and $\theta_D$ are the parameters of the generator and discriminator networks respectively.

However, as the generator in Equation~\eqref{formula:ganloss} is hard to optimise effectively due to the vanishing gradient of the generator when the discriminator reaches the highly confident state with the outputs always equal to 0 initially, it was later improved as:
\begin{equation}
\begin{aligned}\label{formula:wgan_loss}
\mathop{\text{min}}\limits_{\theta_G} 
\mathop{\text{max}}\limits_{\theta_D}
\mathcal{L}(\theta_G, &\theta_D)
=\mathbb{E}_{x \sim p_{\mathrm{data}}(x)}
[\mathop{\text{log}} D_{\theta_D}(x)]\\
&-\mathbb{E}_{z \sim p_{z}(z)}
[\mathop{\text{log}} (D_{\theta_D}(G_{\theta_G}(z)))],
\end{aligned}
\end{equation}
in which the generator can be more easily and stably optimised by the backpropagation algorithm.

With the adversarial min-max training scheme, GAN-based models are adept at reconstructing the fine and natural textures in the generative data. 

\subsection{Basics of Attention and Transformers}

Transformers have a dominant position in the natural language process field and recently have impacted the computer vision (CV) field.
Before Transformers, CNNs had achieved state-of-the-art in many CV tasks, since their deep architectures enlarge the receptive fields progressively and enable CNNs to capture the hierarchies of structured image representations as semantics. However, the convolution, as a basic operator in CNNs, is locally sensitive and lacks long-range dependency.

The receptive field of CNNs is limited by the convolutional kernel and the network depth. Oversized convolutional kernel brings huge computational cost, and overly-deep network depth can cause gradient vanishing. Compared to their counterpart CNNs, Transformers capture long-term dependency with their stacked self-attention blocks.

ViT~\cite{Dosovitskiy2020vit} was proposed for image classification, adopting the encoder in vanilla Transformer with a multi-layer perception (MLP) head for classification. In ViT, input images are converted into patch sequences coupled with position encoding; then compute attention embedding by several cascaded self-attention blocks in the transformer encoder; finally, these attention embeddings are used for classification by an MLP head. 

The core operation in Transformers is the self-attention, which is based on the \emph{query}, \emph{key} and \emph{value} that are derived from the input patch sequence. The self-attention for input $X$ can be expressed as follows:
\begin{equation}
\begin{aligned}\label{formula:kqv}
Q &=X P_{Q}, \quad K=X P_{K}, \quad V=X P_{V},\\
Z &=  \operatorname{Attention}(Q, K, V) \notag \\ &=\operatorname{SoftMax}\left(Q K^{T} / \sqrt{d}+B\right) V,
\end{aligned}   
\end{equation}
in which $Q$, $K$, $V$ denote the query, key, value, and $P_{Q}$, $P_{K}$, $P_{V}$ are corresponding linear projection. $B$ is learnable relative positional encoding and $d$ is the dimension of value and used for scaring here. Such self-attention mechanism calculation is performed for $h$ times and concatenated for multi-head self-attention (MSA).

The self-attention blocks can be expressed as follows:
\begin{equation}
\begin{aligned}\label{formula:selfattention}
&X^{\prime}={\operatorname{MSA}}({\operatorname{LN}}(X))+X,\\
&X^{\prime\prime}={\operatorname{MLP}}({\operatorname{LN}}(X^{\prime}))+X^{\prime},
\end{aligned}
\end{equation}
in which $X$ and $X^{\prime\prime}$ are the input and output of self-attention. ${\operatorname{LN}}(\cdot)$ denotes the Layer Normalisation layer. 

Although ViT-based Transformers have shown their superiority due to their long-range dependency, they suffer from huge computational cost derived from self-attention, especially for high-resolution images. To reduce the computational cost, Liu \textit{et~al.}~\cite{Liu2021swin} proposed the Swin Transformer, adding a special constrict (shifting windows) for self-attention. 


\section{Data Driven Methods for Fast MRI}
\label{Data_Driven_Fast_MRI}

Here we discuss data driven models from the perspective of model structure and training strategies.

\subsection{Model Structure}

Data driven based fast MRI methods mainly fall into 2 main categories: Non-unrolling and Unrolling optimisation.

\subsubsection{Non-Unrolling based Data Driven Methods} 

Most existing data driven fast MRI methods directly learn a mapping from undersampled zero-filled images $x_u$ (or undersampled \textit{k}-space data) to reconstructed images $\hat x_u$, where such mapping is built on a deep network, e.g, CNNs~\cite{Yang2018, Lee2018} and Transformers~\cite{Feng2021_1, Huang2022swinmr} , which can be presented as Equation~\eqref{Non_Unrolling_1}.

Inspired by ResNet~\cite{he2016deep}, most recent non-unrolling based methods applied residual connection and turned the reconstruction function $f(x_u | \theta)$ to a refinement function, which can be presented as Equation~\eqref{Non_Unrolling_2}. The residual connection between the input and output is able to stable the network training and accelerate the convergence:
\begin{equation}
\hat x_u = f(x_u | \theta), \text{ s.t.  } x_u = \mathfrak{F}^{-1} y,
\label{Non_Unrolling_1}
\end{equation}
\begin{equation}
\hat x_u = f(x_u | \theta) + x_u, \text{ s.t.  } x_u = \mathfrak{F}^{-1} y,
\label{Non_Unrolling_2}
\end{equation}
in which $f(\cdot| \theta)$ denotes the deep network parameterised by $\theta$. 

For CNN-based deep networks $f(\cdot| \theta)$, Yang \textit{et~al.}~\cite{Yang2018} applied a modified CNN-based U-Net as the deep network. Lee \textit{et~al.}~\cite{Lee2018} trained two CNN-based residual U-Net for magnitude and phase reconstruction.


For Transformer based deep networks $f(\cdot| \theta)$, Feng \textit{et~al.}~\cite{Feng2021_1} introduced an end-to-end joint reconstruction and super-resolution networks and further advanced the model for these dual tasks by incorporating the model with task-specific novel cross-attention modules.
Huang \textit{et~al.}~\cite{Huang2022swinmr} proposed a swin Transformer based deep network for high-resolution fast MRI.

\subsubsection{Unrolling based Data Driven Methods}

Compared to the non-unrolling networks, the unrolling-based networks are designed to mimic the iterative reconstruction algorithms by iterative blocks, in which all free parameters and functions can be learned through training. 
Most unrolling networks for fast MRI can be divided into two categories. 

The first class was designed to unroll the iterative algorithms~\cite{yang2017admm}, e.g., alternating direction method of multipliers (ADMM), gradient descent (GD), primal-dual (PD) to solve a general CS model given in Equation~\eqref{CS}.
Basic-ADMM-Net~\cite{yang2017admm} was firstly proposed to unroll the ADMM algorithm by designing the iterative blocks where a number of handcrafted parameters  are substituted by the learnable blocks in the network, which combines the advantages in the model-based approach and deep learning approach. Following it, a bunch of methods has emerged to unfold different algorithms for solve different models. 

The second, and more popular approach is to use the image that is generated by the non-unrolling network unit $f(x_u|\theta)$ as a reference image for regularisation~\cite{schlemper2017deep}. The formulation of this method can be described as
\begin{equation}\label{formula:unroll_denoise}
\argmin_{x \in X} \frac{1}{2}|| \mathfrak{M}(x) - y ||_2^2 +  \lambda \| x - f(x_u|\theta) \|_2^2
\end{equation}
Integrating the solution to $\eqref{formula:unroll_denoise}$ and the CNN unit, the iterative structure can be formulated as:
\begin{equation}
\left\{\begin{array}{l}
x^{(l)} = \left(1+\lambda \mathfrak{M}^{\mathrm{T}} \mathfrak{M}\right)^{-1}\left(z^{(l-1)}+\lambda \mathfrak{M}^{\mathrm{T}} f\right),\\
z^{(l)}=f(x^{(l)}|\theta),
\end{array}\right.
\end{equation}
in which $f(x_u|\theta)$ can be selected as CNN-based or Transformer-based structure.

Unrolling the iterative algorithms with more flexible parameters, the unrolling based models are able to learn a general end-to-end mapping with better interpretable structures compared to other non-unrolling networks.

\subsection{Training Strategy}

\subsubsection{Supervised Learning}

Most of the supervised fast MRI methods is trained by paired ground truth images $x_t$ and reconstructed images $\hat x_u$ through a content loss function as follows:
\begin{equation}
\begin{aligned}\label{formula:loss_cont_sup}
\mathop{\text{min}}\limits_{\theta} 
\mathcal{L}_{\mathrm{cont}}({\theta}) =
\mathcal{L}(g(x_t), g(\hat x_u)),
\end{aligned}
\end{equation}
in which $g(\cdot)$ is a mapping to the different domain where the distance function $\mathcal{L}(\cdot)$ is performed. 

The mapping $g(\cdot)$ are usually selected among the following equations:
\begin{equation}\label{formula:g}
g(x)=\left\{\begin{aligned}
&x,~&\text{for pixel-wise loss}, \\
&\mathfrak{F} x,~&\text{for frequency loss}, \\
&f_\text{nn}(x),~&\text{for perceptual loss},
\end{aligned}\right.
\end{equation}
in which $\mathfrak{F}$ denotes the Fourier transform matrix and $f_\text{nn}(\cdot)$ denotes a pre-trained deep neural network, e.g., VGG-16~\cite{Simonyan2015vgg}. 

The distance function $\mathcal{L}(\cdot)$ is typically selected among the following equations: 

\begin{equation}\label{formula:L}
\mathcal{L}(x, \hat x)=\left\{\begin{aligned}
&\| x - \hat x \|_1,~~~~~~~~~~\text{L1 Loss}, \\
&\| x - \hat x \|_2^2,~~~~~~~~~~\text{L2 Loss}, \\
&\sqrt{\| x - \hat x \|^2_2 + \epsilon^2},~\text{Charbonnier Loss},
\end{aligned}\right.
\end{equation}
in which $\epsilon$ is a constant.

For GAN-based fast MRI, adversarial loss is also applied for training:
\begin{equation}
\begin{aligned}\label{formula:gan_loss_sup}
\mathop{\text{min}}\limits_{\theta_G} 
\mathop{\text{max}}\limits_{\theta_D}
\mathcal{L}_{\mathrm{adv}}(\theta_G, &\theta_D)
=\mathbb{E}_{{x_t} \sim p_{\mathrm{t}}({x_t})}
[\mathop{\text{log}} D_{\theta_D}({x_t})]\\
&-\mathbb{E}_{{x_u} \sim p_G({x_u})}
[\mathop{\text{log}} (D_{\theta_D}({G_{\theta_G}(x_u)})].\\
\end{aligned}
\end{equation}

In many fast CS-MRI works, the training loss is designed as a linear combination of several different loss functions:
\begin{equation}\label{formula:Loss_total}
\mathcal{L}_{\mathrm{tot}} = \sum_i \alpha_{i} \mathcal{L}_{i}
\end{equation}

\subsubsection{Unsupervised Learning}

The challenge for supervised learning is the requirement of the large high-quality dataset as ground truth data. To solve this problem, many unsupervised learning based fast MRI methods have been proposed recently.

Most the unsupervised learning based fast MRI methods attempt to minimise the difference between the \textit{k}-space information of reconstructed images and the undersampled measurement $y$ at the undersampled \textit{k}-space locations~\cite{chen2022ai}, which can be expressed as follows:

\begin{equation}\label{formula:loss_cont_unsup}
\mathop{\text{min}}\limits_{\theta} 
\mathcal{L}_{\mathrm{cont}}({\theta}) =
\mathcal{L}(g(y), g(\mathfrak{M}(\hat x_u))).
\end{equation}

Similarly, the adversarial loss introduced in Equation~\ref{formula:gan_loss_sup} for unsupervised learning is transformed into 
\begin{equation}
\begin{aligned}\label{formula:gan_loss_unsup}
\mathop{\text{min}}\limits_{\theta_G} 
\mathop{\text{max}}\limits_{\theta_D}
\mathcal{L}_{\mathrm{adv}}(\theta_G, &\theta_D)
=\mathbb{E}_{{y} \sim p({y})}
[\mathop{\text{log}} D_{\theta_D}({y})]\\
&-\mathbb{E}_{{y} \sim p({y})}
[\mathop{\text{log}} (D_{\theta_D}(\mathfrak{M}({G_{\theta_G}(y)}))].
\end{aligned}
\end{equation}



Guided by the above mentioned losses, deep networks can capture a great deal of image statistics by their structure even without ground truth data~\cite{Lempitsky2018dip} and also reserve the reality of the constructed images.



\section{MR Physics for Data Driven Models}

\subsection{Coupling with Parallel Imaging}

There are two types of parallel imaging methods: Image-domain techniques based on the sensitivity encoding (SENSE) method, as well as \textit{k}-space approaches based on simultaneous acquisition of spatial harmonics (SMASH) and generalised autocalibrating partial parallel acquisition (GRAPPA). 

The SENSE technique was used in traditional parallel imaging in image space, and it has two distinguishing approaches: (1) After applying an inverse Fourier transform to the picture, the aliasing artefacts are then removed in image space. (2) precomputed, explicit coil sensitivity maps are created using either a distinct reference scan or a completely sampled block of data at the centre of \textit{k}-space to get information on receive coil sensitivities. 

GRAPPA, which employs linear shift-invariant convolutional kernels to interpolate missing \textit{k}-space lines using uniformly-spaced acquired \textit{k}-space lines, is the most clinically utilised \textit{k}-space reconstruction approach for parallel imaging. The convolutional kernels are calculated for each subject, similar to the coil sensitivity estimation in SENSE-type reconstruction, from either a single reference scan or a fully-sampled block of data at the centre of \textit{k}-space, termed the autocalibrating signal (ACS). The fully-sampled acquisition locations defined by the kernel size, as well as the associated missing entries, are identified using a sliding window technique in this calibration region

GRAPPA is a linear approach that suffers from noise amplification due to coil shape and acceleration rate, despite its widespread usage in clinical practice. As a result, numerous solutions for reducing noise in reconstruction have been presented in the literature. SPIRiT stands for iterative self-consistent parallel imaging reconstruction. It uses correlations between nearby \textit{k}-space points to enforce self-consistency among the \textit{k}-space data in different receiver coils. SPIRiT estimates a linear shift-invariant convolutional kernel from ACS in the same way as GRAPPA does. In GRAPPA, this convolutional kernel estimated a missing \textit{k}-space point using information from acquired lines in a neighbourhood. The kernel in SPIRiT contains contributions from all points in a neighbourhood surrounding a particular \textit{k}-space point, both acquired and absent, across all coils. According to the self-consistency principle, the whole \textit{k}-space data should stay intact after this convolution. The SPIRiT objective function additionally contains a term that ensures data consistency, where undersampling can be done with any pattern, including random patterns that are commonly used in compressed sensing.

The use of deep neural networks to enhance \textit{k}-space interpolation techniques utilising non-linear approaches in a data driven way has recently piqued interest. Based on how the interpolation functions are trained, these techniques may be classified into two classes. The first group trains deep neural networks for interpolation using scan-specific ACS lines, comparable to known interpolation methods such as GRAPPA. The second group employs training datasets, which are comparable to the deep learning techniques working in the image space (cf. Equation~\eqref{Non_Unrolling_1}). For example, RAKI~\cite{Akcakaya2019}, similar to GRAPPA, used a scan-specific CNN trained on \textit{k}-space centre lines to interpolate missing lines in \textit{k}-space from a fully-sampled \textit{k}-space centre. GrappaNet~\cite{sriram2020grappanet} simplified the reconstruction by using traditional imaging methods, followed by a fine-tuning step. More recently proposed methods also tried to improve the estimation of coil sensitivity functions from limited ACS in SENSE-based reconstruction~\cite{peng2022deepsense}, and also incorporate GAN based models~\cite{lv2021pic}. A dedicated review of parallel imaging using deep learning can also be found in Knoll \textit{et~al.}~\cite{knoll2020deep}.

\subsection{Simultaneous Multi-Slice Imaging}

Another fast MRI approach is simultaneous multi-slice (SMS), also known as multi-band (MB). MRI slices are generally obtained one at a time in a sequential manner in standard 2D acquisitions, which can result in extended scan times if a high number of slices are required. The RF pulses used to acquire the slices, on the other hand, can be configured (with varied frequencies) to pick several slices using just a single RF pulse. This would normally result in images from each slice overlaid on top of each other, rendering them useless for clinical purposes. Similar to parallel imaging, having multiple coils with differing sensitivities allows image reconstruction techniques to be utilised to separate the slices in SMS imaging.

SMS has comparable benefits to CS based fast MRI. There are, nevertheless, certain distinct limits. Residual aliasing artefacts, when the aliasing comes from other slices, are one type of artefact that can appear in SMS acquired images. This is also known as slice leakage. Moreover, SMS is limited in several MRI sequences in clinical practice because of the high specific absorption ratio. SMS has been enhanced by a few technical developments, e.g., Controlled Aliasing In Parallel Imaging Results In Higher Acceleration (CAIPIRINHA), which allows the slices sampled to be closer together. Another advancement, blipped CAIPIRINHA, has allowed SMS to be used to echo planer imaging, greatly increasing the applicability of the SMS. More importantly, SMS should not degrade SNR, unlike parallel imaging, because the data is completely sampled.

Recent development has been manifested by coupling the SMS and deep learning based data driven methods. For instance, Le \textit{et~al.}~\cite{le2021deep} developed an end-to-end deep learning solution for fast SMS based myocardial perfusion reconstruction. Li \textit{et~al.}~\cite{li2022simultaneous} designed a U-Net based CNN to process and separate complex multi-slice overlapping-echo signals that realised a reliable SMS $\text{T}_2$ mapping.

\subsection{Optimisation of Sampling Trajectory and Patterns}

\begin{figure*}[thpb]
  \centering
  \includegraphics[scale=0.35]{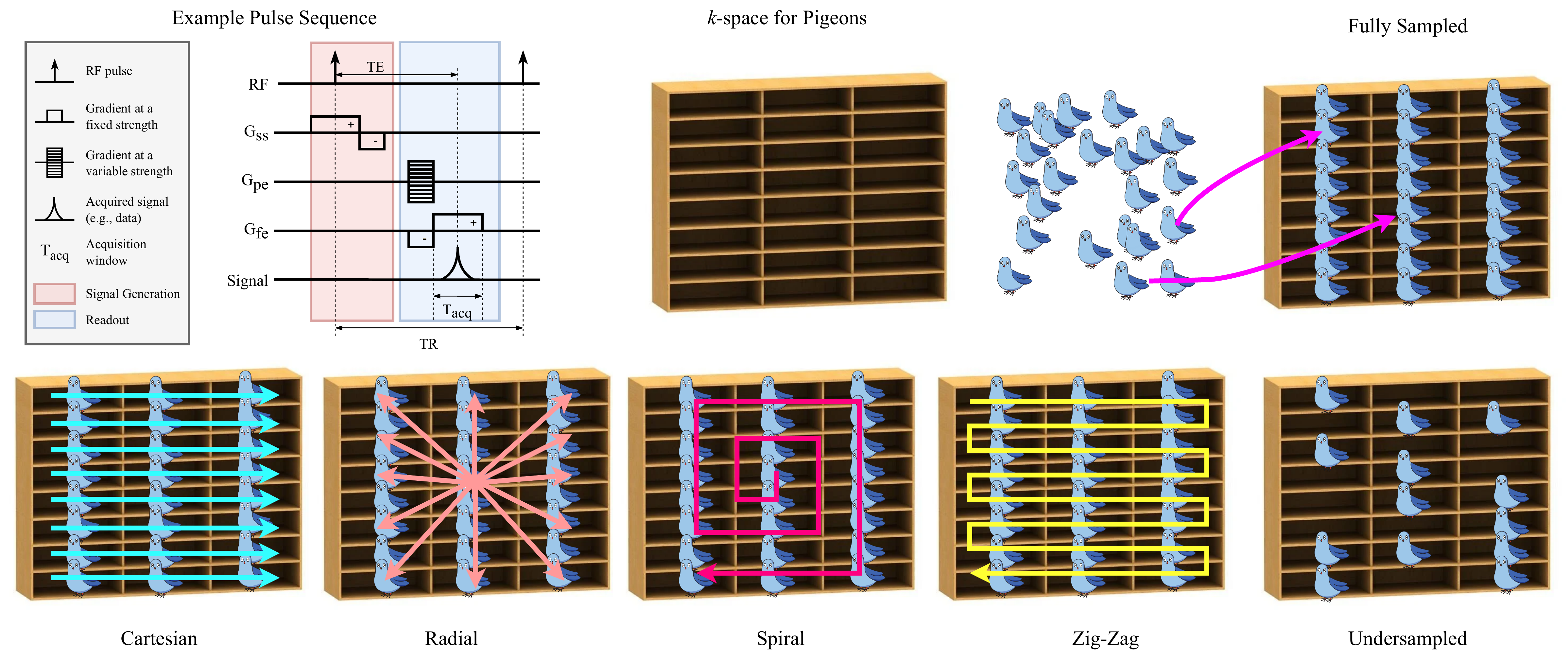}
  \caption{
    (Top Left) The scanner activities are controlled by the pulse sequence. In this example, there are five horizontal lines, representing (i) radiofrequency (RF) signal for excitation, (ii) activities of the slice selection gradient ($\text{G}_\text{ss}$), (iii) activities of the phase encoding gradient ($\text{G}_\text{pe}$), (iv) activities of the frequency encoding gradient ($\text{G}_\text{fe}$), and the signal reception. Conceptually, the pulse sequence can be viewed as consisting of modular blocks (light red/blue boxes). In this example, there are two modules, one for signal creation and the other for signal readout. The scanner is supposed to repeat this timeline, and the period of each iteration is called repetition time (TR). TE: Time to Echo; $\text{T}_\text{acq}$: duration of the signal acquisition window. (Top Right) Analogy of putting pigeons into pigeonholes as acquiring data in the \textit{k}-space and the fully-sampled \textit{k}-space. (Bottom) Four commonly used \textit{k}-space trajectories and the undersampled \textit{k}-space. It is of note that not all undersampling patterns can fit the pulse sequence design.   
  }
  \label{fig:2}
\end{figure*}

Deep learning based data driven fast MRI methods have sprung up across many different scanning sequences, but limitations of clinical translation still abound (see Section~\ref{unrealistic_simulation} for more details). Most previous research has concentrated on developing better fast MRI models given a pre-determined acquisition trajectory, disregarding the issue of trajectory optimisation (Figure \ref{fig:2}). Weiss \textit{et~al.}~\cite{weiss2021} proposed a physics informed learned optimised trajectories (PILOT) method for fast MRI, in which the MR scanner hardware acquisition parameters and constraints are incorporated into the learning pipeline to model the \textit{k}-space trajectories alongside the image reconstruction network optimisation. Although methods similar to PILOT can help point the way toward surmounting the difficulties of on-scanner implementation and clinical translation, these methods entail greedy algorithms or reinforcement learning that require more computational resources and cause training instability. 

\section{Example Applications and Open Datasets}

To facilitate efficient development and fair comparison of the deep learning based data driven fast MRI, we introduce a few widely used open access data in this section as summarized in Table \ref{tab:dataset}.

\begin{table}[htbp]
  \centering
  \caption{A summary of open-access MRI dataset and supported tasks.}
  \scalebox{0.70}{
    \begin{tabular}{ccccccc}
    \toprule
    \multirow{2}[2]{*}{DATASET} & \multirow{2}[2]{*}{Anatomy} & \multicolumn{4}{c}{Tasks}     & \multirow{2}[2]{*}{qMRI} \\
          &       & Reconstruction & Classification & Segmentation & Detection &  \\
    \midrule
    IXI   & Brain & \cmark     & \xmark      & \xmark      & \xmark      & \cmark \\
    fastMRI(+) & Brain, Knee & \cmark     & \cmark     & \xmark      & \cmark     & \xmark \\
    SKM-TEA & Knee  & \cmark     & \cmark     & \cmark     & \cmark     & \cmark \\
    OCMR  & Cardiac & \cmark     & \xmark      & \xmark      & \xmark      & \xmark \\
    \bottomrule
    \end{tabular}}%
  \label{tab:dataset}%
\end{table}%

\subsection{Neuroimaging (e.g., IXI Dataset and fastMRI)}

IXI Dataset\footnote{https://brain-development.org/ixi-dataset/} is an open-access dataset that provides 600 MR images from normal, healthy subjects. The MR image acquisition protocol for each subject includes T1, T2 and PD-weighted images, magnetic resonance angiography (MRA) images and diffusion-weighted images (15 directions). The IXI dataset is comprised of NII files that follow the NIFTI format. 

FastMRI\footnote{https://fastmri.med.nyu.edu/}~\cite{Knoll2021fastmri} is an open-access dataset that provides raw and DICOM data from MRI acquisitions of knees and brains (DICOM data is non-overlapping with the raw data).
FastMRI+\footnote{https://github.com/microsoft/fastmri-plus}~\cite{Zhao2021fastmriplus}, as an updated version, further provides corresponding bounding box annotations and pathology category labels.
The raw brain MRI data is comprised of 6970 multi-coil fully-sampled brain MRI scans. For each scan, axial T1 weighted, T2 weighted and FLAIR images are included (some of the T1 weighted acquisitions with admissions of contrast agent).
In addition to the brain raw data above, fastMRI provides DICOM brain MRI data, comprising 10,000 axial 2D images volumes. For each DICOM data, T1 weighted, T2 weighted and T2 FLAIR images are included.

\subsection{Body MRI (e.g., fastMRI and SKM-TEA Dataset)}

In fastMRI, the raw knee MRI data is comprised of 1594 multi-coil fully-sampled knee MRI scans, including coronal proton-density weighted with (PDFS, 798 scans) and without (PD, 796 scans) fat suppression.
In addition to the knee raw data above, fastMRI provides 10000 DICOM knee MRI data. Each DICOM data consists of coronal PDFS and PD, sagittal PD, sagittal T2 weighted with fat suppression (T2FS) and axial T2FS.

Stanford Knee MRI with Multi-Task Evaluation dataset (SKM-TEA)\footnote{https://github.com/StanfordMIMI/skm-tea}~\cite{Desai2022skmtea} is an open-access dataset that provides a quantitative MRI (qMRI) knee data for multi-task evaluation. The SKM-TEA dataset is comprised of 155 clinical MRI scans. For each scan, the raw \textit{k}-space and DICOM image data, with corresponding T2 qMRI DICOM parameter maps, tissue segmentation maps and localised pathology labels are provided.

\subsection{Cardiac MRI (e.g., OCMR Dataset)}

Open-access Multi-coil \textit{k}-space Dataset for Cardiovascular Magnetic Resonance Imaging dataset (OCMR)\footnote{https://ocmr.info/}~\cite{Chen2020ocmr} is an open-access dataset that provides multi-coil \textit{k}-space data for cardiac cine. The OCMR dataset is comprised of HDF5 files that follow the ISMRMRD format. 
There are 74 fully-sampled scans (comprising 183 slices) and 212 free-breathing prospectively undersampled scans (comprising 842 slices) for cardiac cine in the latest version of the OCMR dataset (October 08, 2020).

\section{Common Pitfalls}

Previous studies have shown several benefits of deep neural network based techniques over conventional constrained reconstruction using specified regularisers. First, instead of using predefined and fixed sparsifying transforms, deep learning based methods have been tuned to a specific MRI reconstruction job and can therefore achieve superior residual artefact removal. This is especially important when the sample trajectory utilised does not meet the incoherence criteria of CS, which is frequently the case with clinical parallel imaging methods. Second, deep learning frameworks have disentangled the computationally intensive training process from the inference step. It is crucial in MRI reconstruction to have diagnostic-quality images accessible immediately after the scan so that technicians and radiologists may decide if a certain sequence needs to be repeated or acquisition settings need to be adjusted. Conventional CS based methods have still suffered from slow iterative reconstruction while deep learning based inference models can be performed on specialised GPUs, on the other hand, are more efficient. 

When compared to traditional CS based methods, current data driven deep learning reconstruction frameworks have a number of limitations. First, they require the availability of a representative curated training data set so that the trained model may generalise to unseen testing data. Recent techniques in the literature have either been trained using hundreds of instances rather than millions, as is usual in deep learning for computer vision, or trained on synthetic non-medical data that is publicly available from existing databases. However, there remain issues that may limit the development and deployment of data driven deep learning reconstruction for specific applications. For example, due to spatio-temporal resolution limits, some applications in imaging of moving organs, such as the heart, or imaging of brain connections, such as diffusion tensor imaging, cannot be captured with fully-sampled data. More importantly, MRI scans with contrast injection, e.g., late gadolinium enhancement (LGE) cardiac magnetic resonance imaging for atrial fibrillation patients, must be acquired within a certain time frame before complete contrast wash-out. Such physical or physiological constraints make it difficult to obtain fully-sampled training labels on such datasets, emphasising the importance of scan-specific techniques or unsupervised training procedures.

During the training stage, these reconstruction approaches also demand significant computational resources. Because of the growing availability of GPUs, this could be fewer concerns. However, the requirements of expertise in deploying and maintaining such GPU powered workstations with data driven deep learning methods can stonewall the prevalence of those methods, especially in remote sites without such infrastructures or human resources. The availability of on-demand cloud-based deep learning services can be a possible solution, but concerns about clinical data privacy may raise ethical problems.

A more serious problem is that unlike traditional CS based fast MRI, deep learning models are typically non-convex. Their features, particularly in terms of failure mechanisms and generalisation potential for everyday clinical usage, are less well understood compared to traditional iterative techniques based on convex optimisation. It was recently demonstrated, for example, that while reconstructions generalise well in terms of changes in MRI image contrast across training and test data, they are susceptible to systematic variations in SNR. It is also unclear how specific the trained models must be. There is an open question if it is sufficient to train a single model for all sorts of MR tests, or do distinct models for various anatomical locations, pulse sequences, acquisition trajectories, and acceleration factors, as well as scanner manufacturers, field strengths, and receiving coils, need to be trained. While pre-training a large number of separate models for different MRI sequences is feasible in clinical practice, if certain models do not generalise concerning scan parameter settings that are typically tailored to the specific anatomy of an individual patient by the MR technologist, this will have a significant impact on their translational potential and, ultimately, their clinical use.

\begin{figure}[thpb]
  \centering
  \includegraphics[scale=0.65]{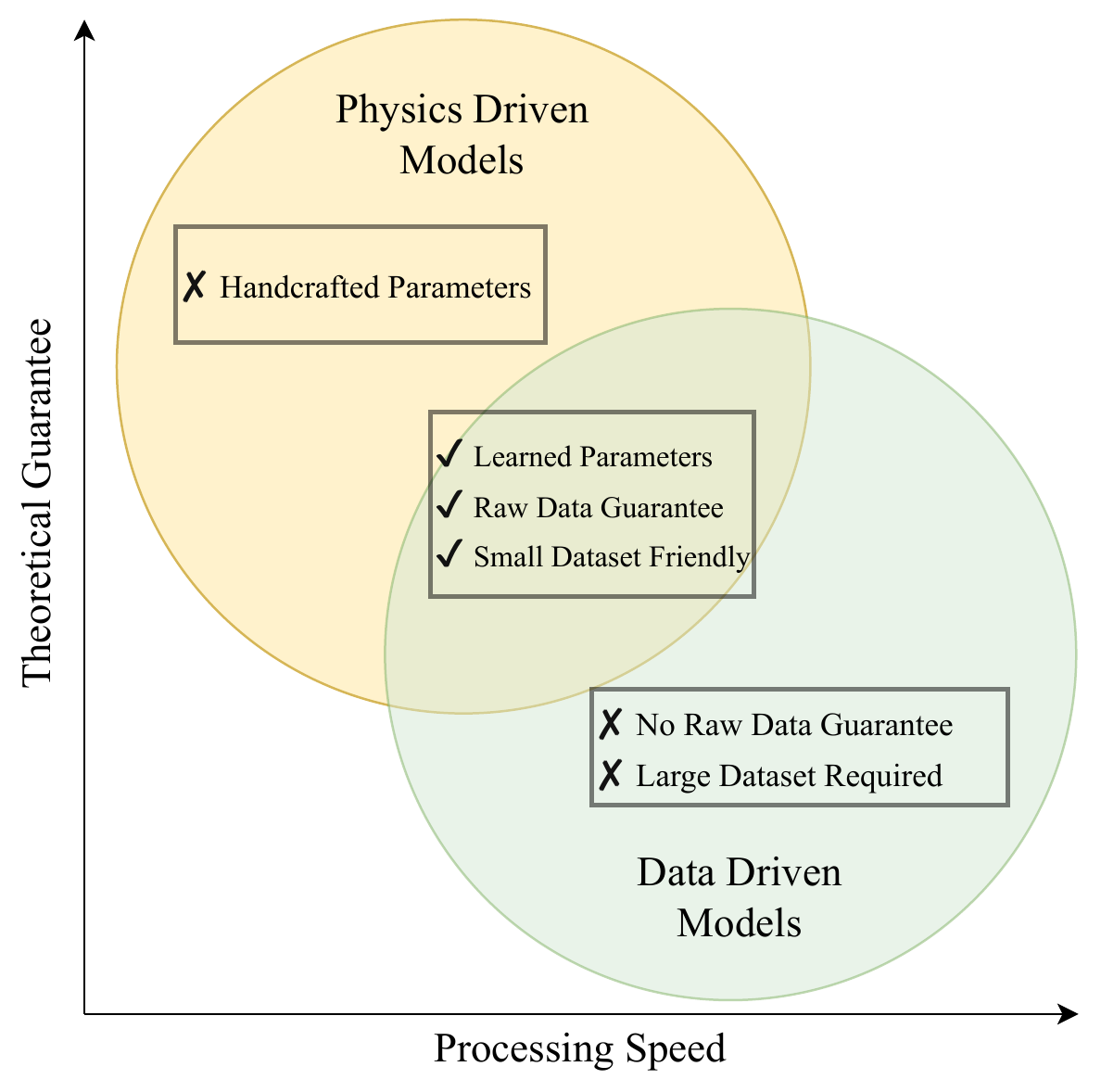}
  \caption{
    Advantages of synergistic data and physics driven models.
  }
  \label{fig:3}
\end{figure}

The selection of regularisation parameters is a practical difficulty when using regularisation, e.g., in conventional CS based fast MRI. An image quality metric is required for automatic parameter optimisation. The mean-square error to a reference \emph{ground truth} has historically been employed, and numerous additional metrics, such as structural similarity, have been devised, but none of the available methodologies adequately represents a human observer's subjective judgement of the image quality. This is especially significant in the current data drive deep learning based reconstruction algorithms, which frequently need the estimation of thousands or even millions of parameters. Even more crucial than apparent visual quality, the reconstruction must preserve or increase diagnostic accuracy. It is significantly more difficult to demonstrate this, and it is currently done by thorough testing at each particular institution whenever a new approach is presented. Below we summarised the three most prominent limitations in current data driven deep learning based fast MRI and the advantages of synergistic data and physics driven models are outlined in Figure \ref{fig:3}.

\subsection{Instabilities with Perturbations}

\begin{figure*}[thpb]
  \centering
  \includegraphics[scale=0.45]{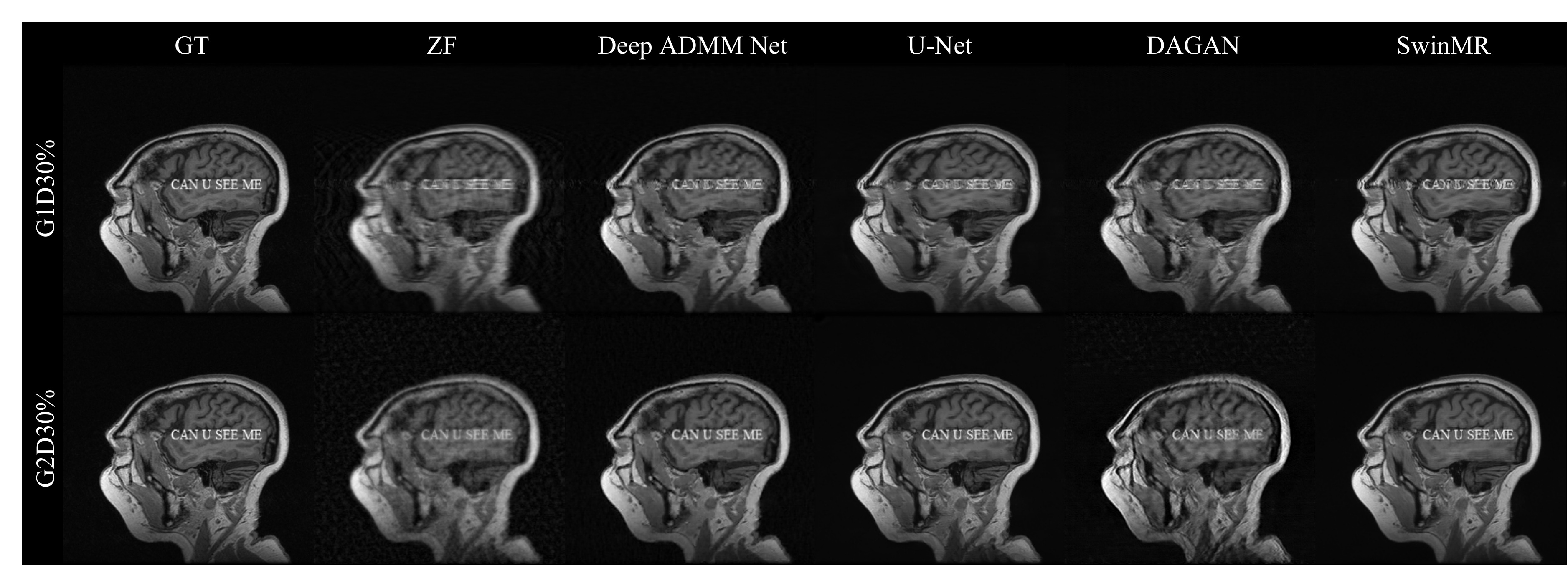}
  \caption{
    The reconstruction results of different data driven models with structured perturbations. From left to right, the ground truth image, and the reconstruction results of zero-filled (ZF), Deep ADMM Net, U-Net, DAGAN, and SwinMR.
  }
  \label{fig:4}
\end{figure*}

Despite deep learning based data driven methods having excelled in fast MRI tasks, research studies have also shown instabilities and the generalisability and reproducibility can be poor. Antun \textit{et~al.}~\cite{antun2020instabilities} published a study investigating the instabilities for deep based reconstruction. In particular, three staggering findings are (1) Small even hard to be detected perturbations, either in the image or sampling domain, may cause severe artefacts in the reconstruction; (2) an insignificant structural change, such as a tumour, may not be captured in the reconstructed image; and (3) more training samples may result in poorer performance, which is counter-intuitive. Because the instability phenomenon is not easily remedied, we recommend future studies carry out such instability tests on top of the normal quantitative evaluations (Figure \ref{fig:4}).   

\subsection{Unrealistic Simulation and Delayed Clinical Translation} \label{unrealistic_simulation}

Most existing fast MRI studies are based on simulation, e.g., following a supervised learning paradigm to undersample the original \textit{k}-space measurement (ground truth) and try to learn the recovery function between retrospectively downsampled signal and \emph{fully sampled ground truth}. However, the effects of the unrealistic simulation are not immaterial and can delay the clinical translation of the developed deep learning based data driven fast MRI. 

For example, currently, most 2D fast MRI studies rely on 1D random undersampling to generate a sampling pattern that follows a 1D Gaussian distribution. This results in a higher amount of undersampling in the high frequency regions of the \textit{k}-space while retaining low frequencies to preserve the overall image structure. Because sampling along the frequency encoding direction is quick, and only the phase encoding direction restricts the acquisition time, 1D undersampling should be simulated instead of using 2D undersampling. When it comes to 3D reconstruction, 2D Gaussian and Poisson disc masks are frequently utilised to speed up phase and slice encoding. 

Moreover, two other major problems hinder the clinical translation: (1) The currently used fully sampled ground truth is only nominal. Nowadays, most routinely used MRI sequences are performed with parallel imaging using multi-coils scanning. The algorithms trained on this kind of retrospectively undersampled data will face problems to be implemented on the scanners by simply mimicking parallel imaging based reconstruction rather than incorporating parallel imaging into the reconstruction chain. (2) Most simulated undersampling patterns are not realistic or feasible to be implemented on the scanner, e.g., widely used Gaussian undersampling could not be implemented using MRI sequence programming, which is far from being a straightforward clinical application (Figure \ref{fig:1}).

\subsection{Problems of the Current Evaluation Metrics}

Current evaluation metrics can be mainly categorised into fidelity evaluation metrics and perceptual evaluation metrics.

\subsubsection{Fidelity Evaluation Metrics}

The most commonly used fidelity evaluation metrics for existing fast MRI research are Peak Signal-to-Noise Ratio (PSNR) and Structural Similarity Index (SSIM). 

PSNR is the ratio between maximum signal power and noise power, measuring the fidelity of the representation, which is defined as follows:
\begin{equation}
\begin{aligned}\label{formula:psnr}
\operatorname{PSNR}(x, {\hat x})
=10 \cdot \log _{10}(\frac{L^{2}}{\frac{1}{N} \sum_{i=1}^{N}(x(i)-{\hat x}(i))^{2}}),
\end{aligned}
\end{equation}
in which $x$ and $\hat x$ are two paired images. $L$ and $N$ denote the data range and the pixel number of $x$ and $\hat x$.

SSIM quantifies the structural similarity between two images based on luminance, contrast, and structures, which is defined as follows:
\begin{equation}
\begin{aligned}\label{formula:ssim}
\operatorname{SSIM}(x, {\hat x})
=\frac{2 \mu_{x} \mu_{{\hat x}}+\kappa_{1}}{\mu_{x}^{2}+\mu_{{\hat x}}^{2}+\kappa_{1}} \cdot \frac{2 \sigma_{x {\hat x}}+\kappa_{2}}{\sigma_{x}^{2}+\sigma_{{\hat x}}^{2}+\kappa_{2}},
\end{aligned}
\end{equation}
in which $x$ and $\hat x$ are two paired images. $\mu$ and $\sigma$ denote the mean and standard deviation. $\sigma_{x {\hat x}}$ is the co-variance between $x$ and $\hat x$. $\kappa_{1}$ and $\kappa_{2}$ are constant relaxation terms.

Both fidelity metrics are based on simple and shallow functions and direct comparisons between paired images. Although these evaluation metrics provide a relatively accurate result of the fidelity of images, they failed to reflect the visual quality for human observers~\cite{Zhang2018lpips}. For example, blurring tends to have a larger influence on perceptual evaluation metrics than on PSNR.

\subsubsection{Perceptual Evaluation Metrics}

Fr\'echet Inception Distance (FID)~\cite{Heusel2017fid}, as a deep learning based perceptual assessment, measures the Fr\'echet distance between two multivariate Gaussian distribution $\mathcal{N}(\mu, \sigma)$. 
Two sets of images are converted into distributions of deep representations by a pre-trained Inception-V3~\cite{Szegedy2016inceptionv3} model, then the FID between these two distributions is defined as follows:
\begin{equation}
\begin{aligned}\label{formula:fid}
\mathrm{FID}(\mathbb{I}_r,\mathbb{I}_g)&=\|\mu_{r}-\mu_{g}\|^{2} \\ &+\operatorname{Tr}(\sigma_{r}+\sigma_{g}-2(\sigma_{r} \sigma_{g})^{\frac{1}{2}}),
\end{aligned}
\end{equation}
in which $\mathbb{I}_r$ and $\mathbb{I}_g$ are the multivariate distributions converted from reconstructed and ground truth image sets. A lower FID indicates a more perceptual result.

Learned Perceptual Image Patch Similarity (LPIPS)~\cite{Zhang2018lpips} measures image perceptual while reducing manual intervention. 
LPIPS first converts reconstructed $x_r$ and ground truth $x_g$ images to their deep embeddings $y_r^l$ and $y_g^l$ for layer $l$ by a deep network, then scares the deep embedding by vector $w_{l}$, computes the L2 distance and finally computes average spatially and sum layer-wisely. LPIPS is defined as follows:
\begin{equation}
\begin{aligned}\label{formula:lpips}
d(x_{r}, x_{g})=\sum_{l} \frac{1}{H_{l} W_{l}} \sum_{h, w}\|w_{l} \odot(\hat{y}_{r h w}^{l}-\hat{y}_{g h w}^{l})\|_{2}^{2}.
\end{aligned}
\end{equation}
in which $H_{l}$ and $W_{l}$ are height and width of the feature map for layer $l$. $\odot$ denotes pixel-wise multiplication.
A lower LPIPS indicates a more perceptual result.

FID and LPIPS are based on the comparison of deep representation, which correlates well with visual quality for human observers. 
However, perceptual evaluation metrics are insensitive to minor differences, especially those details with \emph{same style but different pixel-wisely}.
Specifically, in the reconstruction task with a low sampling rate, some models may provide the reconstructed images with `rich but wrong' details. 
Phenomena can be observed that these `rich but wrong' details lead to a worse PSNR and SSIM, but a better FID ~\cite{Huang2022swinmr}.

In addition, the Mean Opinion Score (MOS) is a subjective perceptual evaluation metric, which asks experienced radiologists to rate the reconstructed images. 
Typically, Likert scales from 1 (poor), 2 (fair), 3 (good) to 4 (very good) are applied for rating based on the reconstruction quality. 
Although MOS is a reliable perceptual quality assessment, it suffers from inter-/inner-raters bias, the variance of rating criteria, the time-consuming rating process.

\section{Important Considerations and Future Perspectives}

\subsection{The Importance of Data Harmonisation}

MRI suffers from high variability across vendors  and acquisition protocols. The non-biological heterogeneity caused by different brands of devices (manufacturer, magnetic field, field strength, etc.), acquisition protocols (voxel size, echo time, etc.) greatly obstructs the studies that rely on multi-source datasets. Different from the CT imaging that can follow some guidelines during the data collection, there does not exist a general guideline for MRI since its intensity value has no physical meaning. 

To address this issue, data harmonisation is proposed to unite multi-centre datasets and alleviate the non-biological heterogeneity through computational strategies (e.g., machine learning and image processing)~\cite{nan2022data}. The harmonisation of MR scans mainly includes two schemes, the feature-wise and the sample-wise. The feature-wise approaches align the cohort distribution through statistical strategies during the feature extraction, while the sample-wise approaches harmonise the raw image scans through image synthesis or image processing. Studies have shown that the non-biological variances can be effectively reduced by implementing either feature-wise or sample-wise harmonisation strategies.

\subsection{The Importance of Explainability}

Clinicians remain wary of using deep learning based methods in medical image analysis including data driven fast MRI. This can be ascribed to the lack of explainability of these deep learning models. Explainable artificial intelligence (XAI) is a new topic in AI that tries to give rationale, transparency, and traceability of frequently black-box deep learning algorithms, as well as testability of causal assumptions. In biomedical signal and image processing, especially applications in digital healthcare, determining causation is especially important to justify why a decision is taken and why one intervention or treatment option is preferred over others. In deep learning, XAI is a step toward realising the FATE (Fairness, Accountability, Transparency, and Ethics) and FAIR (Findable, Accessible, Interoperable, Reusable) principles. 

However, developing XAI in biomedical signal and image processing is difficult. Here, we primarily address it from the angles of AI technique customisation, nonlinear data, problem-solving difficulty, and learning bias. To begin with, cutting-edge AI approaches have not been created for biomedical data directly. They come from computer vision, image recognition, automated reasoning, cognition, and even statistics. It might be difficult to describe how existing AI approaches can be used in biomedical data research. Instead of merely applying AI algorithms to unique datasets, they should be adapted or even changed for optimal performance and interpretation. However, because there is no established AI theory to guide it and the needed degree of explainability varies with different application areas, such a customisation process may not be readily accomplished in a short amount of time.

There are currently limited studies on the explainability of fast MRI. Two possible future directions may be explored: (1) Schlemper \textit{et~al.}~\cite{schlemper2018stochastic} proposed a stochastic deep CS method for the quantification and visualisation of the reconstruction uncertainty. Such uncertainty measurements could provide explainability of the fast MRI networks (e.g., highlight regions the algorithm may be incompetent) that may also enlighten further downstream clinical tasks. More importantly, how to use the uncertainty measurements to further reinforce the performance of the fast MRI, especially with more constraints imposed with MR physics will be certainly an open area to delve into. (2) Combining fast MRI with downstream segmentation or classification tasks to demonstrate the salience maps, class activation maps, or attention maps~\cite{yang2022unbox}, which can be indirect indicators of the explainability of the reconstruction network.

\subsection{Federated Learning}

Similar to other emerging deep learning based approaches in medical data processing, data driven fast MRI also relies on huge volumes of collected data. Existing MRI data is not completely utilised by deep learning, partly because it is stored in data silos and access to this data is restricted due to privacy concerns. However, without appropriate data, data driven fast MRI will not be able to attain its full potential and, eventually, will not be able to close the gap from research to clinical practice.

In general digital healthcare settings, federated learning enables multi-centre and multi-scanner studies across different sites to develop accurate and robust deep learning models without revealing or exchanging the underlying data, embedding privacy-by-design into the model and solving crucial concerns such as data privacy, security, and access rights. Federated learning allows for the collective acquisition of insights, such as in the form of a consensus model, without exposing patient data beyond the firewalls of the institutions/hospitals in which they are housed. Instead, deep learning takes place locally at each participating institution/hospital, with just model properties (e.g., parameters, gradients) shared. Current efforts to increase federated learning privacy often draw on existing classical cryptographic protocols such as secure multiparty computation and differential privacy~\cite{li2020federated}.

Federated learning for fast MRI is a new topic and only a few pilot studies have been published, e.g.,~\cite{guo2021multi}. However, federated learning based fast MRI will not only realise privacy-preserved learning but will also enhance the generalisation of the developed data driven fast MRI. This is because essentially most widely used routinely acquired MRI are not quantitative, and federated learning can certainly improve the robustness of the algorithms by incorporating the multi-centre and multi-scanner data. General speaking, future research on federated learning based fast MRI will deal with (1) data diversity --- MRI data is inherently complex and varied. Factors such as acquisition disparities, scanner differences, and area demographics all influence the data variety. Furthermore, data may not be dispersed equitably across participating institutions; (2) System architecture --- a federated learning system that permits access to MRI data housed by many institutions/nodes must accommodate for data integrity, encryption, and varying computational resources, e.g., weaving together synchronous and asynchronous operations to avoid performance penalties caused by the straggler effect; (3) Traceability --- federated learning may also impose additional opacity without reaching the MRI data directly. 


\subsection{Novel Fast Multi-Parametric Quantitative MRI}

Essentially, fast MRI including conventional CS and data driven based deep learning methods can perform better when more data redundancy exists. On top of structural MRI, which is the gold standard imaging technique for many clinical diagnosis and monitoring problems, multi-parametric quantitative MRI (qMRI) can quantify tissue parameters, allowing for the identification of microstructural processes associated with tissue remodelling in ageing or pathological cases. Besides, multi-parametric qMRI allows for the investigation of physiologically distinct microstructural processes that may precede changes in tissue structure. This opens new avenues for a more thorough characterisation of tissue modifications by identifying early deterioration of microstructural integrity as well as particular disease-related patterns. We anticipate more development and trends in data driven based fast MRI in multi-parametric qMRI, e.g., methods for MR fingerprinting, diffusion tensor imaging, and MR Spectroscopy.

\section{Conclusion}

The fast emergence of deep learning in computer vision has resulted in significant changes in the landscape of deep learning based data driven fast MRI approaches over the last few years. Earlier deep learning based fast MRI approaches has witnessed a burgeoning interest in novel deep neural network designs such as the data consistency layer, variational network, GAN, residual learning, cross-domain networks, attention and newly proposed transformer based models. Recently, the research community has realised that more MR physics constraints must be imposed to ensure the smooth translation from simulation based studies to MR scanners for the readiness of clinical applications. However, with the expanding diversity of deep learning based fast MRI algorithms, it is still difficult to identify an adequate and fair comparison standard. Nonetheless, we feel that the thrill of this topic stems not only from advancing beyond benchmark works but also from developing new benchmarks for undiscovered fast MRI applications, e.g., exploring the untapped potential of multi-parametric qMRI. Both open access MRI data repository and the development of federated learning can enhance the generalisability and reproducibility of the deep learning based data driven fast MRI. With the advances of XAI, we envision a brilliant future of fast MRI, ushering in a new era of personalised medicine in digital healthcare, with the advent of high throughput and low-cost scanning, as well as more reliable quantitative imaging biomarkers extraction and explainable analysis.


%




\ifCLASSOPTIONcaptionsoff
  \newpage
\fi



%


\bibliographystyle{IEEEtran}
\bibliography{References}

%








\end{document}